\documentclass[amsmath,twocolumn,amssymb,showkeys,superscriptaddress,floatfix,aps,prd,10pt]{revtex4}
\usepackage{graphicx,epstopdf}
\usepackage[caption=false]{subfig}
\usepackage{multirow}
\usepackage{appendix}
\usepackage{xcolor}
\usepackage[colorlinks=true, urlcolor=blue, linkcolor=blue, citecolor=green]{hyperref}

\def\xslash{x\!\!\!\slash }

\begin{document}

\title{Magnetic moment of the $X_1(2900)$ state in the diquark-antidiquark picture}
\author{U.~\"{O}zdem}%
\email[]{ulasozdem@aydin.edu.tr}
\affiliation{ Health Services Vocational School of Higher Education, Istanbul Aydin University, Sefakoy-Kucukcekmece, 34295 Istanbul, T\"{u}rkiye}
\author{K. Azizi}
\email[]{kazem.azizi@ut.ac.ir }
\affiliation{Department of Physics, University of Tehran, North Karegar Avenue, Tehran
14395-547, Iran}
\affiliation{Department of Physics, Do\v{g}u\c{s} University, Dudullu-\"{U}mraniye, 34775
Istanbul, T\"{u}rkiye}

\begin{abstract}
Motivated by the discovery of fully open-flavor tetraquark states $X_0(2900)$ and $X_1(2900)$ by the LHCb Collaboration, the magnetic dipole moment of the $X_1(2900)$ state with the quantum numbers $ J^{P} =  1^{-}$   is  determined in the  diquark-antidiquark picture   using the light-cone sum rules. The numerical result is obtained as $ \mu_{X_1}=0.79^{+0.36}_{-0.39}\,\mu_N$. 
The magnetic moments of hadrons encompasses useful knowledge on the distributions of charge and magnetization their inside, which can be used to better understand their geometrical shapes and quark-gluon organizations.
 The observation of the $X_0(2900)$ and $X_1(2900)$ as the first two fully open-flavor multiquark states has opened a new window  for investigation of the  exotic states. The obtained results in the present  study may shed light on the future experimental and theoretical searches on the properties of  fully open-flavor multiquark states.
\end{abstract}
\keywords{Magnetic dipole moment, $X_1(2900)$ state, Light-cone sum rules}
\maketitle

\section{Introduction}

In 2021, the LHCb Collaboration observed two clear peaks in the
$D^-K^+$ invariant mass spectrum of the $B^+\to D^+D^-K^+$
decay~\cite{LHCb:2020bls,LHCb:2020pxc}.  Their spectroscopic parameters are
measured to be
\begin{eqnarray}
X_0(2900): M&=&2866\pm7\pm2~\text{MeV},\nonumber\\
\Gamma&=&57\pm12\pm4~\text{MeV};\nonumber\\
X_1(2900): M&=&2904\pm5\pm1~\text{MeV},\nonumber\\
\Gamma&=&110\pm11\pm4~\text{MeV}.\nonumber
\end{eqnarray}

The $D^-K^+$  configuration suggests that the  quark constituents of $ X_0(2900) $ and  $ X_1(2900) $ should be
$[\bar{c}\bar{s}][ud]$, which means that they are  fully open-flavor exotic hadrons. The possible spin-parity quantum numbers of $X_0(2900)$ and $X_1(2900)$ are  estimated to be $J^P = 0^+$ and $1^-$, respectively.

This observation triggered interesting phenomenological studies on  these new resonances in the
context of various approaches  and models aiming to elucidate their  nature, quantum numbers and substructure. These studies are assigned different schemes for these particles: Molecular forms of  $\bar{D}^\ast K^\ast$ and $\bar{D}_1K$
interactions in Refs. ~\cite{Chen:2020aos,He:2020btl,Liu:2020nil,Hu:2020mxp,Agaev:2020nrc,Chen:2021tad},
the diquark-antidiquark picture in Refs.~\cite{Chen:2020aos,Karliner:2020vsi,He:2020jna,Wang:2020xyc,Zhang:2020oze,Wang:2020prk,Agaev:2021knl},
and kinematic effects from the triangle
singularities in Refs.~\cite{Liu:2020orv,Burns:2020epm}. The production and
decay properties of these states were investigated in
Refs.~\cite{Huang:2020ptc,Chen:2020eyu,Burns:2020xne,Xiao:2020ltm}, as well.
One can also consult
Refs.~\cite{Albuquerque:2020ugi,Lu:2020qmp,Mutuk:2020igv,Tan:2020cpu,Abreu:2020ony,Qi:2021iyv,Chen:2021erj,Hsiao:2021tyq,Duan:2021bna,Kong:2021ohg,Dong:2020rgs,Bondar:2020eoa}
for other pertinent studies on parameters of these states.

Despite the above experimental and theoretical investigations, their properties remain dubious and determination of their exact nature and substructure  is still problematic. Indeed, the properties of the  $ X_0(2900) $ and  $ X_1(2900) $ tetraquark states have been suggested  differently in different studies. To resolve  these ambiguities, their parameters are needed to be further investigated both in theory and experiment. These studies shall include  complementary investigations of their spectroscopy as well as  their various reactions with other known particles and/or their strong, electromagnetic and weak decay modes.

Inspired by this, in the present  study, we are going to consider the interaction  of $ X_1(2900) $ ($X_1$ for short) tetraquark state with photon and examine the magnetic dipole moment of this state within the framework of the light-cone sum rules (LCSR).  In the LCSR method, a two-point  correlation function is calculated in two different steps. In the first step, it is obtained in terms of hadronic parameters such as form factors, electromagnetic multipole moments, etc., which is called hadronic representation. In the second step, it is calculated in terms of quark-gluon degrees of freedom, which is called the QCD representation. Then, the correlation functions obtained from these two different ways are related to each other using quark-hadron duality assumption. Finally, the Borel transform and continuum subtraction are performed to  suppress the contributions of the possible higher states and continuum. By this way, one obtains the sum rules for the desired physical quantities.   In the calculations we use the distribution amplitudes (DAs) of the on-shell photon. 
When calculating the magnetic dipole moment of $X_1$, we will consider that this state has a diquark-antidiquark configuration. 
There are few studies in the literature where the magnetic dipole moments of the open-flavor exotic states have been investigated, see for instance the Refs.~\cite{Azizi:2018mte,Azizi:2018jky,Azizi:2021aib}.

The paper is organized as follows. In Section \ref{formalism}, some details of the calculations of the magnetic dipole moment of $X_1$ in  LCSR method  is given. In Sec. \ref{numericc}, we present our numerical results and discussions.
  In Section \ref{sumup}, we discuss obtained results and conclude with brief notes.

 \section{ formalism }\label{formalism}

As we have mentioned above, at the beginning of the analytic calculations of the magnetic dipole  moment,  it is necessary to select a sufficient two-point correlation function in the background electromagnetic field, which plays an important role in the LCSR method. It is written as 
\begin{equation}
 \label{edmn01}
\Pi _{\mu \nu }(p,q)=i\int d^{4}xe^{ip\cdot x}\langle 0|\mathcal{T}\{J_{\mu}(x)
J_{\nu }^{ \dagger }(0)\}|0\rangle_{\gamma}, 
\end{equation}%
where $J_{\mu}(x)$ and $\gamma$ represent the interpolating current of $X_{1}$ state and the external electromagnetic field, respectively. We need explicit expression of  $J_{\mu}(x)$  to proceed in the calculations. In the diquark-antidiquark picture, $J_{\mu}(x)$ can be written in the following form
\begin{eqnarray}
J^{X_1}_{\mu}(x)&=&\varepsilon^{abc} \varepsilon^{amn}\big[u^{Tb}(x)C\gamma_5 d^c(x)\big] \nonumber\\
&&\times \big[ \bar{c}^m(x)\gamma_\mu \gamma_5 C \bar{s}^{Tn}(x)\big],
\end{eqnarray}
where $ C $ is the charge conjugation matrix; and  $a, b, c, m, n$ are color indices.

 In the hadronic representation,  two complete sets of the initial and final hadronic states are inserted into the correlation function. By isolating the    lowest $X_{1}$ state contribution, we obtain,
\begin{align}
\label{edmn04}
\Pi_{\mu\nu}^{Had} (p,q) &= {\frac{\langle 0 \mid J_\mu (x) \mid
X_{1}(p, \varepsilon^\theta) \rangle}{p^2 - m_{X_{1}}^2}}\nonumber\\
& \times \langle X_{1}(p, \varepsilon^\theta) \mid X_{1}(p+q, \varepsilon^\delta) \rangle_\gamma \nonumber\\
& \times
\frac{\langle X_{1}(p+q,\varepsilon^\delta) \mid {J_\nu^{ \dagger}} (0) \mid 0 \rangle}{(p+q)^2 - m_{X_{1}}^2} + \cdots,
\end{align}
where  dots denote the effects of the higher states and continuum. The matrix elements in Eq. (\ref{edmn04}) are expressed as
\begin{align}
\label{edmn05}
&\langle X_{1}(p+q,\varepsilon^\delta) \mid {J_\nu^{ \dagger}} (0) \mid 0 \rangle=\lambda_{X_{1}} \varepsilon_\nu^\delta\,,
\\
\nonumber\\
&\langle 0 \mid J_\mu(x) \mid X_{1}(p,\varepsilon^\theta) \rangle = \lambda_{X_{1}} \varepsilon_\mu^\theta\,,
\\
\nonumber\\
&\langle X_{1}(p,\varepsilon^\theta) \mid  X_{1} (p+q,\varepsilon^{\delta})\rangle_\gamma = - \varepsilon^\tau (\varepsilon^{\theta})^\alpha (\varepsilon^{\delta})^\beta \Big\{ G_1(Q^2)\nonumber\\
& \times (2p+q)_\tau ~g_{\alpha\beta}  + G_2(Q^2)  g_{\tau\beta}~ q_\alpha -  g_{\tau\alpha}~ q_\beta) \nonumber\\ &- \frac{1}{2 m_{X_{1}}^2} G_3(Q^2)~ (2p+q)_\tau q_\alpha q_\beta  \Big\},\label{edmn06}
\end{align}
where $\lambda_{X_{1}}$ is the residue of $ X_1 $; and  $ \varepsilon_\mu^\theta\ $, $ \varepsilon_\nu^\delta\ $ and   $ \varepsilon^\tau $ are  the initial and final  polarization vectors of the $X_{1}$ and photon polarization vector, respectively. Here,  
$G_1(Q^2)$, $G_2(Q^2)$ and $G_3(Q^2)$ are electromagnetic form factors,  with  $Q^2=-q^2$.

Using Eqs. (\ref{edmn04})-(\ref{edmn06}) and after performing some necessary calculations,  the final form of the correlation function is  obtained as
\begin{align}
\label{edmn09}
 &\Pi_{\mu\nu}^{Had}(p,q) =  \frac{\varepsilon_\rho \, \lambda_{X_{1}}^2}{ [m_{X_{1}}^2 - (p+q)^2][m_{X_{1}}^2 - p^2]}
 \Big\{ G_2 (Q^2) \nonumber\\
 & \times \Big(q_\mu g_{\rho\nu} - q_\nu g_{\rho\mu} -
\frac{p_\nu}{m_{X_{1}}^2}  \big(q_\mu p_\rho - \frac{1}{2}
Q^2 g_{\mu\rho}\big) 
  \nonumber\\
 &  +
\frac{(p+q)_\mu}{m_{X_{1}}^2}  \big(q_\nu (p+q)_\rho+ \frac{1}{2}
Q^2 g_{\nu\rho}\big)
-  
\frac{(p+q)_\mu p_\nu p_\rho}{m_{X_{1}}^4} \, Q^2
\Big)
\nonumber\\
&
+\mbox{other independent structures}\Big\}\,+\cdots.
\end{align}

To calculate the magnetic dipole moment, we need  to calculate only the form factor  $G_2(Q^2)$, which is called the magnetic form factor,
\begin{align}
\label{edmn07}
&F_M(Q^2) = G_2(Q^2)\,.
\end{align} 
 At static limit,  $Q^2 = 0 $,   $F_M(Q^2 = 0)$ is proportional to the
 magnetic dipole moment $\mu_{X_{1}}$ for real photon:
\begin{align}
\label{edmn008}
&\mu_{X_{1}} = \frac{ e}{2\, m_{X_{1}}} \,F_M(Q^2 = 0).
\end{align}

The correlation function, on the other hand,  is determined in terms of the QCD degrees of freedom and the photon distribution amplitudes in the second window called the QCD side. In the QCD representation, we use the Wick's theorem to contract the corresponding quark fields to get the correlation function in terms of the quark propagators and DAs of the photon.  After replacing the explicit expression of the interpolating current in the correlation function and applying the Wick's theorem,  we get
\begin{eqnarray}
\Pi _{\mu \nu }^{\mathrm{QCD}}(p,q)&=&i\int d^{4}xe^{ipx}\varepsilon^{abc} \varepsilon^{amn} \varepsilon^{a^{\prime }b^{\prime }c^{\prime }} \varepsilon^{a^{\prime }m^{\prime }n^{\prime }} \nonumber\\
&& \times
\mathrm{Tr}[ \gamma _{5}\widetilde{S}_{u}^{bb^{\prime
}}(x)\gamma _{5}S_{d}^{cc^{\prime }}(x)] 
\mathrm{Tr}[ \gamma _{\mu }\gamma _{5}\widetilde{S}%
_{s}^{n^{\prime }n}(-x)\nonumber\\
&& \times
 \gamma _{5}\gamma _{\nu }S_{c}^{m^{\prime }m}(-x)%
] \mid 0 \rangle_{\gamma},\nonumber\\  \label{edmn11}
\end{eqnarray}%
where%
\begin{equation*}
\widetilde{S}_{c(q)}^{ij}(x)=CS_{c(q)}^{ij\mathrm{T}}(x)C,
\end{equation*}%
with $S_{q(c)}(x)$ being the quark propagators.
In the $x$-space for the light-quark propagator we use 
\begin{align}
\label{edmn12}
S_{q}(x)&=S_q^{free}
- \frac{\langle \bar qq \rangle }{12} \Big(1-i\frac{m_{q} \xslash}{4}   \Big)
- \frac{ \langle \bar qq \rangle }{192}m_0^2 x^2  \Big(1\nonumber\\
&-i\frac{m_{q} \xslash}{6}   \Big)
-\frac {i g_s }{32 \pi^2 x^2} ~G^{\mu \nu} (x) \Big[\rlap/{x} 
\sigma_{\mu \nu} +  \sigma_{\mu \nu} \rlap/{x}
 \Big],
\end{align}
where,
\begin{align}
 S_q^{free}=i \frac{{\xslash}}{2\pi ^{2}x^{4}} -\frac{m_q}{4\pi ^{2}x^{2}}  .
 \end{align}
The charm-quark propagator is given, in terms of  the second kind Bessel functions $K_{i}(x)$, as
\begin{align}
\label{edmn13}
S_{c}(x)&=S_c^{free}
-\frac{g_{s}m_{c}}{16\pi ^{2}} \int_0^1 dv\, G^{\mu \nu }(vx)\Bigg[ \big(\sigma _{\mu \nu }{\xslash}
  +{\xslash}\sigma _{\mu \nu }\big) \nonumber\\
  & \times
  \frac{K_{1}\Big( m_{c}\sqrt{-x^{2}}\Big) }{\sqrt{-x^{2}}} 
+2\sigma_{\mu \nu }K_{0}\Big( m_{c}\sqrt{-x^{2}}\Big)\Bigg].
\end{align}%
where,
\begin{align}
 S_c^{free}=\frac{m_{c}^{2}}{4 \pi^{2}} \Bigg[ \frac{K_{1}\Big(m_{c}\sqrt{-x^{2}}\Big) }{\sqrt{-x^{2}}}
+i\frac{{\xslash}~K_{2}\Big( m_{c}\sqrt{-x^{2}}\Big)}
{(\sqrt{-x^{2}})^{2}}\Bigg] .
 \end{align}

The correlation function in QCD representation includes two different contributions: Perturbative and non-perturbative.
Practically, the perturbative contribution, in which the photon interacts with one of the quarks perturbatively, can be computed by the replacing  one of the light or c-quark propagators by
\begin{align}\label{lightprop}
S^{free} \rightarrow \int d^4z\, S^{free} (x-z)\,\rlap/{\!A}(z)\, S^{free} (z)\,,
\end{align}
 and the other  three propagators with their perturbative or free parts. In the last equation  we also use the perturbative parts of the propagators in right hand side as is seen. 

For the non-perturbative contribution, in which the photon is radiated at long distances, the correlation function can be computed by replacing one of the light quark propagators by 
\begin{align}\label{pertquarkprop}
S_{\alpha\beta}^{ab} \rightarrow -\frac{1}{4} (\bar{q}^a \Gamma_i q^b)(\Gamma_i)_{\alpha\beta},
\end{align}
where $\Gamma_i = I, \gamma_5, \gamma_\mu, i\gamma_5 \gamma_\mu, \sigma_{\mu\nu}/2$,  
and  the remaining light and heavy propagators with their full expressions. We perform all the possible permutations in the perturbative and non-perturbative parts of the correlation function. 
 When Eq.~(\ref{pertquarkprop}) is employed in computation of the non-perturbative effects, we observe that matrix elements of the forms  $\langle \gamma(q) | \bar{q}(x) \Gamma_i q(0) | 0\rangle$ and $\langle \gamma(q) | \bar{q}(x) \Gamma_i G_{\mu\nu}q(0)  | 0\rangle$ appear. These matrix elements are written in terms of  the photon distribution amplitudes (see Ref. \cite{Ball:2002ps}).  Using these matrix elements in terms of photon's DAs and the expressions of the propagators given above, the QCD representation of the correlation function in coordinate space is obtained. We perform Fourier transformation to carry the calculations to the momentum space.

The LCSR for the magnetic dipole  moment of $X_1$ state can be acquired by matching the functions $\Pi_{\mu\nu}^{QCD} (p)$ and $\Pi_{\mu\nu}^{Had} (p)$ from both the QCD and hadronic sides. We choose the $\varepsilon_\nu q_\mu$ structure and equate the coefficients of this structure from both sides to each other. We apply a double Borel transformation with respect to   -$p^2$ and -$(p + q)^2$ and also continuum subtraction procedure based on the standard prescriptions of the method in order to suppress the contributions of the higher states and continuum (fore details see for instance  Refs. \cite{Azizi:2018duk, Agaev:2016srl}).  As a results, we get the desired LCSR for the magnetic dipole moment as
\begin{align}
 &\mu_{X_{1}}  = \frac {e^{\frac{m_{X_{1}}^2}{M^2}}}{\lambda_{X_{1}}^2 } \,\, \Delta (M^2,s_0),
\end{align}
where the explicit expression of $\Delta (M^2,s_0)$ function, which represents the final form of the  QCD side of the calculations, is  presented in the Appendix. In obtaining the last result, we use $ \frac {1}{M^2} =\frac {1}{M_1^2}+\frac {1}{M_2^2}$ with $ M_1^2 $ and $ M_2^2 $ being the Borel parameters in the initial and final channels, respectively. We set $ M_1^2= M_2^2 =2 M^2 $  as the initial and final particles are the same. We will fix $ M^2 $ and continuum threshold ($ s_0 $), coming from the continuum subtraction procedure and appearing inside the $\Delta (M^2,s_0)$ function,  based on the standard criteria of the method in next section.



\section{Numerical analysis}\label{numericc}

In this section, we present our numerical prediction for the magnetic dipole moment of the $X_1$ state.  
To get our numerical results, the main input parameters are the photon distribution amplitudes. The distribution amplitudes of the photon, which depend on various non-perturbative parameters are borrowed from Ref.~\cite{Ball:2002ps}. 
We use the values of some other  input parameters as follow: $m_u=m_d=0$, $m_s =96^{+8}_{-4}\,\mbox{MeV}$,
$m_c = (1.275\pm 0.025)\,$GeV, $ m_{X_1}= 2904\pm5\pm1~\text{MeV}$,    
$f_{3\gamma}=-0.0039~GeV^2$~\cite{Ball:2002ps},  
$\langle \bar ss\rangle $= $0.8 \langle \bar uu\rangle$ with 
$\langle \bar uu\rangle $=$(-0.24\pm0.01)^3\,$GeV$^3$~\cite{Ioffe:2005ym},  
$m_0^{2} = 0.8 \pm 0.1$~GeV$^2$~\cite{Ioffe:2005ym}, $\langle \frac{\alpha_s}{\pi} G^2 \rangle =(0.012\pm0.004)$ $~\mathrm{GeV}^4 $~\cite{Belyaev:1982cd} and 
$\lambda_{X_{1}}= m_{X_1} f_{X_1} $ with  $f_{X_1}= (2.1 \pm 0.4) \times 10^{-3}$~GeV$^4$~\cite{Agaev:2021knl}.

In the LCSR method, in addition to the DAs and  above parameters, there are two extra auxiliary parameters as mentioned before: The  Borel mass parameter $M^2$ and the continuum threshold $s_0$. The continuum threshold is not totally arbitrary but it shows the energy scale at which, the excited states and continuum begin to contribute to the correlation function. 
The physical quantities like the magnetic dipole moment  are expected to be independent of these helping parameters. In practice, however, there appear some residual dependence on these parameters which are entered as the uncertainties to the presented results. To find the working intervals of these auxiliary parameters,  we demand that  both the continuum and higher states contributions have to be sufficiently suppressed and series of the operator product expansion (OPE) in QCD side converge. To this end, in technique language, we define the pole contribution (PC) as
\begin{align}
 \mbox{PC} =\frac{\Delta (M^2,s_0)}{\Delta (M^2,\infty)},
 \end{align}
 and require that it should exceed $20 \%$ of the total contributions, which is typical for the multiquark states. We also demand that the  series of light-cone expansion converges and contributions of the  higher twist and higher condensate terms are less than $10 \%$ of the total contribution. 
These considerations  lead to the following working windows for $M^2$ and $s_0$:
\begin{eqnarray}
 3.0~\mbox{GeV}^2 \leq M^2 \leq 3.5~\mbox{GeV}^2 \nonumber\\
  11.0~\mbox{GeV}^2 \leq s_0 \leq 12.5~\mbox{GeV}^2. \nonumber
\end{eqnarray}
Our numerical analyses show that, by considering these working regions for the auxiliary parameters, for the magnetic dipole moment of the $X_1$ state $  PC$ varies  within  the interval $30\%\leq PC \leq 64\%$ corresponding to the upper and lower limits of the Borel mass parameter. 
When we analyze the OPE convergence, we see that  the contribution of the higher twist and higher dimensional terms in OPE is  $2 \%$ of the total and the series show a good convergence. It is worth mentioning that the above interval for the continuum threshold corresponds to $s_0 \simeq (m_{X_1}+0.5^{+0.1}_{-0.1})^2$~GeV$^{2}$, which is typical in hadronic spectrum.
Therefore, the chosen working windows for $M^2$ and $s_0$ well fulfill the requirements of the light-cone sum rules method.

\begin{figure}[htp]
\centering
\includegraphics[width=0.48\textwidth]{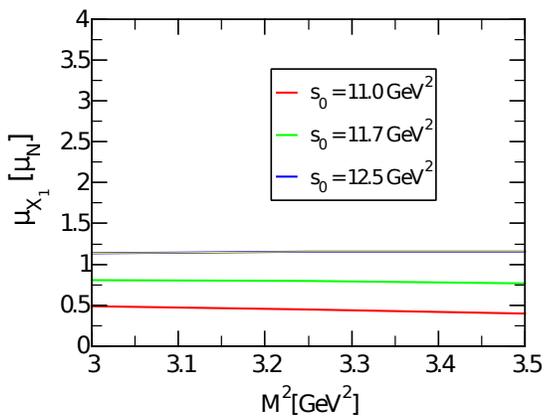}
 \caption{The magnetic dipole moment of the $X_1$ state  versus $M^2$ at three fixed values of $s_0$.}
 \label{Msqfig1}
  \end{figure}

In Fig.\ref{Msqfig1}, we depict the dependence of the magnetic dipole moment of the $X_1$ state on $M^2$ at three fixed values of $s_0$. As one can see from this figure, the magnetic dipole moment show a good stability with respect  to the  variation of the Borel mass parameter. Although the dependence on   $s_0$ is  considerable,  however,  it remains within the limits allowed by the method to calculate the magnetic dipole moment.

 
Considering all the input parameters, DAs of the photon and the working intervals of auxiliary parameters, our prediction for the magnetic dipole moment of the $X_1$ state, both in its natural unit  ($\frac{ e}{2\, m_{X_1}}$) and nuclear magneton ($\mu_N=\frac{ e}{2\, m_{N}}$), is
\begin{align}
 \mu_{X_1}= 2.43^{+1.13}_{-1.21}\frac{ e}{2\, m_{X_1}}=0.79^{+0.36}_{-0.39}\,\mu_N.
\end{align}
The order of  magnitude  for the magnetic dipole moment  shows that  measurement of $  \mu_{X_1} $ is accessible in the future  experiments.

\section{Summary and Conclusion}\label{sumup}

Motivated by the discovery of fully open-flavor tetraquark states $X_0(2900)$ and $X_1(2900)$ by the LHCb Collaboration, the magnetic dipole moment of the $X_1(2900)$ tetraquark state have been determined using the light-cone  sum  rules assigning  the diquark-antidiquark structure with the quantum numbers $ J^{P} =  1^{-}$ for this state. 
The magnetic moments of hadrons encompasses useful knowledge about the distribution of charge and magnetization their inside, which helps us to understand their nature, quark-gluon organization and geometrical shape.
The existing theoretical predictions on the spectroscopic parameters of $X_1(2900)$ tetraquark  and their comparison with the experimental  data have also given rise to different assignments on the internal structure of this state. Calculations of electromagnetic parameters of the exotic states like their magnetic dipole moment can be useful in establishing the nature of these states.
The observation of the $X_0(2900)$ and $X_1(2900)$ by LHCb as the first two fully open-flavor multiquark states has opened a new platform for investigation of the exotic states. More experimental and theoretical research is required to fully understand the properties of this class of hadrons. The magnitude obtained for the magnetic dipole moment of  $X_1(2900)$  shows a possibility to measure it in future experiments. The obtained result in the present study may be useful for analyses of the future experimental  data on parameters of fully open-flavor multiquark states.

\section*{ACKNOWLEDGMENTS}
K. Azizi is thankful to Iran Science Elites Federation (Saramadan)
for the partial  financial support provided under the grant number ISEF/M/400150.

 \begin{widetext}
  \appendix
\subsection*{Appendix: Explicit expression of \texorpdfstring{$\Delta (M^2,s_0)$}{}}
 In this appendix, we present the explicit expression of the function $\Delta (M^2,s_0)$ entering into the sum rule for the magnetic moment of the $X_1$ state. It is obtained as
 \begin{align}
 \Delta (M^2,s_0) &= -\frac {e_c} {36864 m_c^2 \pi^6} \Bigg[
   3 m_c^{12} \big (3 I[-6, 4] - 4 I[-5, 3]\big)- 
    48 m_c^{10} \big (I[-5, 4] + I[-4, 3]\big) - 
    48  m_c^9 m_s I[-4, 3]   \nonumber\\
    &+ 
    3 m_c^8 \Big (30 I[-4, 4]+ P_ 1 (I[-4, 2] - 2 I[-3, 1]) + 
       96 m_s P_ 3 \pi^2 \big (I[-4, 2] - 2 I[-3, 1]\big) - 
       24 I[-3, 3]\Big) + 
     \nonumber\\
    &144  m_c^7 \big (4 P_ 3 \pi^2 I[-3, 2] - m_s I[-3, 3]\big)- 
    12 m_c^6 \Big (-32 m_ 0^2 m_s P_ 3 \pi^2 I[-3, 1] + 
       512 P_ 2^2 \pi^4 I[-3, 1] + P_1 I[-3, 2]  \nonumber\\
    &+ 
       96 m_s P_ 3 \pi^2 I[-3, 2] + 6 I[-3, 4] + P_ 1 I[-2, 1]+ 
       96 m_s P_ 3 \pi^2 I[-2, 1] + 4 I[-2, 3]\Big)  \nonumber\\
    &- 
    12  m_c^5 \Big (48 P_ 3 \pi^2 (m_ 0^2 I[-2, 1]+ 2 I[-2, 2]) + 
       m_s (P_ 1 I[-2, 1] + 12 I[-2, 3])\Big) + 
    3 m_c^4 \Big (64 m_ 0^2 m_s P_ 3 \pi^2 I[-2, 1] \nonumber\\
    &- 
       1024 P_ 2^2 \pi^4 I[-2, 1] + 3 P_ 1 I[-2, 2] + 
       288 m_s P_ 3 \pi^2 I[-2, 2] + 7 I[-2, 4] - 2 P_ 1 I[-1, 1]- 
       192 m_s P_ 3 \pi^2 I[-1, 1]  \nonumber\\
    &- 4 I[-1, 3]\Big) + 
    8 \Big (-192 P_ 2^2 \pi^4 (m_ 0^2 I[0, 0] + 2 I[0, 1]) + 
       m_s P_ 3 \pi^2 \big (P_ 1 I[0, 0] + 
           24 m_ 0^2 I[0, 1]\big)\Big) \nonumber\\
    &- 192 m_c^2 I[0, 3] - 
    16  m_c \Big (P_ 1 P_ 3 \pi^2 I[0, 0] + 
       24 m_s \big (-8 P_ 2^2 \pi^4 I[0, 0] + I[0, 3]\big)\Big) + 
    12  m_c^3 \big (48 P_ 3 \pi^2 \big (I[-1, 2] \nonumber\\
    & - 
           m_ 0^2 I[1, 0]\big)- 
        m_s \big (4 I[-1, 3] + P_ 1 I[1, 0]\big)\Big)\Bigg]
        \nonumber\\
    &+\frac {e_d} {442368 m_c^2 \pi^6}\Bigg[
   f_ {3\gamma} \pi^2 \Bigg (-11  P_ 1  \big (m_c^6 I[-3, 1] + 
          3 m_c^4 I[-2, 1] + 2 I[0, 1]\big) I_ 1[\mathcal {A}] + 
       12  \Big (5 m_c^{10} I[-5, 3] \nonumber\\
    &- 12 m_c^8 I[-4, 3] - 
          36 m_c^7 m_s I[-3, 2] + 
          m_c^6 \big (48 m_s P_ 3 \pi^2 I[-3, 1] - 9 I[-3, 3]\big) + 
          24 m_c^5 \big (8 P_ 3 \pi^2 I[-2, 1] \nonumber\\
    &+ 
             3 m_s I[-2, 2]\big) - 
          2 m_c^4 \big (24 m_s P_ 3 \pi^2 I[-2, 1] + 
             5 I[-2, 3]\big) - 36 m_c^3 m_s I[-1, 2] + 
          48 m_ 0^2 m_c P_ 3 \pi^2 I[0, 0] \nonumber\\
    &+ 
          8 m_s P_ 3 \pi^2 \big (m_ 0^2 I[0, 0] - 12 I[0, 1]\big) - 
          20 I[0, 3]\Big) I_ 1[\mathcal {V}]\Bigg) +
    4 \Bigg (1152 m_c^4 P_ 2^2 \pi^4 I_ 4[\mathcal {S}] I[-2, 1] \nonumber\\
    &- 
        6 m_c^5 P_ 1 \big (m_c (2 m_c (m_c + m_s) I[-3, 1] + 
              I[-3, 2]) + (3 m_c + 2 m_s) I[-2, 1]\big) + 
        9 m_c^4 P_ 1 I[-2, 2] - 6 m_c^4 P_ 1 I[-1, 1] \nonumber\\
    &- 
        6 m_c^3 m_s P_ 1 I[-1, 1] + 12 m_c^3 m_s P_ 1 I[0, 0] + 
        16 m_c P_ 1 P_ 3 \pi^2 I[0, 0] + 
        8 m_s P_ 1 P_ 3 \pi I[0, 0] \nonumber\\
    &- 
        1152 m_c m_s P_ 2^2 \pi^4 I_ 4[\mathcal {S}] I[0, 0] - 
        48 m_c m_s P_ 1 I[0, 1] + 
        1152 P_ 2^2 \pi^4 I_ 4[\mathcal {S}] I[0, 1] - 
        144 P_ 2^2 \pi^4  \big (6 m_c^6 I[-3, 1] \nonumber\\
    &+ 4 m_c^4 I[-2, 1] + 
           m_ 0^2 I[0, 0] + 2 I[0, 1]\big) I_ 1[\mathcal {S}] + 
        9 P_ 1 I[0, 2] + 18 m_c^3 m_s P_ 1 I[1, 0] + 
        2 f_ {3\gamma} P_ 1 \pi^2 \big (3 m_c^6 I[-3, 1]\nonumber\\
    & + 
           2 m_c^4 I[-2, 1] + 3 I[0, 1] + 
           2 m_c^3 m_s I[1, 0]\big) I_ 6[\psi_a] + 
        4 f_ {3\gamma} P_ 1 \pi^2 \big (3 m_c^6 I[-3, 1] + 
            2 m_c m_s I[0, 0] - I[0, 1]\big) \psi^a[u_ 0]\Bigg)\Bigg]
            \nonumber\\
    & 
    +\frac {e_u} {442368 m_c^2 \pi^6}\Bigg[
   f_ {3\gamma} \pi^2 \Bigg (12  \Big (5 m_c {10} I[-5, 3] - 
          12 m_c^8 I[-4, 3] - 36 m_c^7 m_s I[-3, 2] + 
          m_c^6 \big (48 m_s P_ 3 \pi^2 I[-3, 1] \nonumber\\
    &- 9 I[-3, 3]\big) + 
          24 m_c^5 \big (8 P_ 3 \pi^2 I[-2, 1] + 
             3 m_s I[-2, 2]\big) - 
          2 m_c^4 \big (24 m_s P_ 3 \pi^2 I[-2, 1] + 
             5 I[-2, 3]\big) \nonumber\\
    & - 36 m_c^3 m_s I[-1, 2] + 
          48 m_ 0^2 m_c P_ 3 \pi^2 I[0, 0] + 
          8 m_s P_ 3 \pi^2 \big (m_ 0^2 I[0, 0] - 12 I[0, 1]\big) - 
          44 I[0, 3]\Big) I_ 2[\mathcal {V}] \nonumber\\
    &+ 
       P_ 1 \Big (11  \big (m_c^6 I[-3, 1] + 3 m_c^4 I[-2, 1] + 
              2 I[0, 1]\big) I_ 2[\mathcal {A}] + 
           8  \big (3 m_c^6 I[-3, 1] + 2 m_c^4 I[-2, 1] + 3 I[0, 1] \nonumber\\
    &+ 
              2 m_c^3 m_s I[1, 0]\big) I_ 6[\psi^a] + 
           16 \big (3 m_c^6 I[-3, 1] + 2 m_c m_s I[0, 0] - 
               I[0, 1]\big) \psi^a[u_ 0])\Bigg)
    - 24 m_c^5 (3 m_c + 2 m_s) P_ 1 I[-2, 1] \nonumber\\
    & - 
    4608 P_ 2^2 \pi^4  \big (m_c^4 I[-2, 1] - m_c m_s I[0, 0] + 
       I[0, 1]\big) I_ 3[\mathcal {S}] + 
    576 P_ 2^2 \pi^4  \big (6 m_c^6 I[-3, 1] + 4 m_c^4 I[-2, 1] + 
       m_ 0^2 I[0, 0] \nonumber\\
    &+ 2 I[0, 1]\big) I_ 2[\mathcal {S}] + 
    4 P_ 1 \Big (m_c^4 (9 I[-2, 2] - 6 I[-1, 1]) + 
        8 m_s P_ 3 \pi^2 I[0, 0] + 
        16 m_c \big (P_ 3 \pi^2 I[0, 0] - 3 m_s I[0, 1]\big)\nonumber\\
    & + 
        9 I[0, 2] - 
        6 m_c^3 m_s (I[-1, 1] - 2 I[0, 0] - 3 I[1, 0])\Big)\Bigg]\nonumber
 \end{align}
 
\begin{align}
 &-\frac {e_s} {221184 m_c^2 \pi^4} \Bigg[
   m_c^{12} (6 I[-6, 4] + 8 I[-5, 3]) - 
    24 m_c {10} (I[-5, 4] - I[-4, 3]) + 
    m_c^8 \big (P1 I[-4, 2] + 30 I[-4, 4] \nonumber\\
    &+ 2 P_1 I[-3, 1] + 
       24 I[-3, 3]\big) + 
    2 m_c^6 \big (512 P_ 2^2 \pi^4 I[-3, 1] - P_ 1 I[-3, 2] - 
       6 I[-3, 4] + P_ 1 I[-2, 1] + 4 I[-2, 3]\big) \nonumber\\
    &+ 
    1024 P_ 2^2 \pi^4 I[0, 1] - 3 P_ 1 I[0, 2] + 64 m_c^2 I[0, 3] + 
    P_ 3 \Bigg (864  \big (m_c^6 I[-3, 2] - 
           m_c^4 I[-2, 2]\big)\mathcal {A}[u_ 0] - 
       432 m_c^2 \big (I_ 4[\mathcal{S}] + I_ 4[\mathcal {T}_1] \nonumber\\
    &+ 
           I_ 4[\mathcal {T}_2] - I_ 4[\mathcal {T}_3] - 
           I_ 4[\mathcal {T}_4] - 
           I_ 4[\mathcal {\tilde S}]\big) \big (m_c^4 I[-3, 2] - 
          2 m_c^2 I[-2, 2] + I[-1, 2]\big) + 
       P_ 1 \Big (23 I_ 4[\mathcal {S}] + 23 I_ 4[\mathcal {T}_1] + 
          23 I_ 4[\mathcal {T}_2]\nonumber\\
    & - 
          12 \big (I_ 4[\mathcal {T}_3] + I_ 4[\mathcal {T}_4] + 
              I_ 4[\mathcal {\tilde S}]\big)\Big) I[0, 0] + 
       24 \Big (36 m_c^4 (m_c^4 I[-4, 2] - 2 m_c^2 I[-3, 2] + 
             I[-2, 2]) - P_ 1 I[0, 0]\Big) 
             I_ 6[h_ {\gamma}]\Bigg)\nonumber\\
    & - 
    f_{3\gamma} \Bigg (144  m_c^8 I_ 1[\mathcal {A}] I[-4, 3] + 
        144  m_c^8 I_ 1[\mathcal {V}] I[-4, 3] + 
        288  m_c^6 I_ 1[\mathcal {A}] I[-3, 3] + 
        288 m_c^6 I_ 1[\mathcal {V}] I[-3, 3] + 
        23  m_c^4 P_ 1 I_ 1[\mathcal {A}] I[-2, 1]\nonumber\\
    & + 
        23  m_c^4 P_ 1 I_ 1[\mathcal {V}] I[-2, 1] + 
        144  m_c^4 I_ 1[\mathcal {A}] I[-2, 3] + 
        144  m_c^4 I_ 1[\mathcal {V}] I[-2, 3] + 
        23  P_ 1 I_ 1[\mathcal {A}] I[0, 1] + 
        23  P_ 1 I_ 1[\mathcal {V}] I[0, 1]\nonumber\\
    & + 
        576  I_ 1[\mathcal {A}] I[0, 3] + 
        576  I_ 1[\mathcal {V}] I[0, 3] + 
        24  \Big (m_c^6 (12 m_c^4 I[-5, 3] + 24 m_c^2 I[-4, 3] + 
              P_ 1 I[-3, 1] + 12 I[-3, 3]) - P_ 1 I[0, 1] \nonumber\\
    &+ 
           48 I[0, 3]\Big) I_ 6[\psi^a] + 
        96  \Big (m_c^6 (12 m_c^4 I[-5, 3] + 24 m_c^2 I[-4, 3] + 
              P_ 1 I[-3, 1] + 12 I[-3, 3]) + P_ 1 I[0, 1] + 
           48 I[0, 3]\Big) I_ 6[\varphi_{\gamma}]
           \nonumber\\
    &+ 
        576 m_c^{10} I[-5, 3] \psi^a[u_ 0] + 
        48 m_c^6 P_ 1 I[-3, 1] \psi^a[u_ 0] - 
        576 m_c^6 I[-3, 3] \psi^a[u_ 0] + 
        48 P_ 1 I[0, 1] \psi^a[u_ 0] + 
        48 \Big (m_c^6 \big (12 m_c^4 I[-5, 3]\nonumber\\
    & + 24 m_c^2 I[-4, 3] + 
               P_ 1 I[-3, 1] + 12 I[-3, 3]\big) + P_ 1 I[0, 1] + 
            48 I[0, 3]\Big) \varphi_{\gamma}[u_ 0]\Bigg)\Bigg],
\end{align}
where $P_1 =\langle g_s^2 G^2\rangle$ is gluon condensate, $P_2 =\langle \bar q q \rangle$ stands for u/d quark condensate, and $P_3 = \langle \bar s s \rangle$ represents the s-quark condensate. 
The functions~$I[n,m]$,~$I_1[\mathcal{A}]$,~$I_2[\mathcal{A}]$,~$I_3[\mathcal{A}]$,~$I_4[\mathcal{A}]$,~$I_5[\mathcal{A}]$ and~$I_6[\mathcal{A}]$    are defined as
\begin{align}
I[n,m]&= \int_{m_c^2}^{s_0} ds \int_{m_c^2}^s dl~ e^{-s/M^2}~\frac{(s-l)^m}{l^n}\nonumber\\
 I_1[\mathcal{A}]&=\int D_{\alpha_i} \int_0^1 dv~ \mathcal{A}(\alpha_{\bar q},\alpha_q,\alpha_g)
 \delta'(\alpha_ q +\bar v \alpha_g-u_0),\nonumber\\
  I_2[\mathcal{A}]&=\int D_{\alpha_i} \int_0^1 dv~ \mathcal{A}(\alpha_{\bar q},\alpha_q,\alpha_g)
 \delta'(\alpha_{\bar q}+ v \alpha_g-u_0),\nonumber\\
  I_3[\mathcal{A}]&=\int_0^1 du~ A(u)\delta'(u-u_0),\nonumber\\
  I_4[\mathcal{A}]&=\int D_{\alpha_i} \int_0^1 dv~ \mathcal{A}(\alpha_{\bar q},\alpha_q,\alpha_g)
 \delta(\alpha_ q +\bar v \alpha_g-u_0),\nonumber\\
   I_5[\mathcal{A}]&=\int D_{\alpha_i} \int_0^1 dv~ \mathcal{A}(\alpha_{\bar q},\alpha_q,\alpha_g)
 \delta(\alpha_{\bar q}+ v \alpha_g-u_0),\nonumber\\
 I_6[\mathcal{A}]&=\int_0^1 du~ A(u),\nonumber
 \end{align}
where $\mathcal{A}$ stands for the corresponding photon DAs and $ D_{\alpha_i}  $ is the measure  defined as
\begin{equation}
\int {\cal D} \alpha_i = \int_0^1 d \alpha_{\bar q} \int_0^1 d
\alpha_q \int_0^1 d \alpha_g \delta(1-\alpha_{\bar
q}-\alpha_q-\alpha_g).\nonumber \\
\end{equation}

\end{widetext}

\bibliography{article}

\end{document}